\documentclass[10pt,twocolumn]{article} 
\usepackage{simpleConference}
\usepackage{times}
\usepackage{graphicx}
\usepackage{amssymb}
\usepackage{hyperref}
\usepackage{url}
\usepackage{xcolor}

\begin{document}

\title{Do Fewer Tiers Mean Fewer Tears? Eliminating Web Stack Components to Improve Interoperability}

\author{Adrian Ramsingh, Jeremy Singer, Phil Trinder \\
%\\
%Technical Report \\
University of Glasgow \\
%\today
%\\
%\\
firstname.lastname@glasgow.ac.uk  \\
}

\maketitle
\thispagestyle{empty}

\begin{abstract}

Web applications are structured as  multi-tier stacks of components. Each component may be written in a different language and interoperate using a variety of protocols. Such interoperation  increases developer effort, can introduce security vulnerabilities, may reduce performance and require additional resources. A range of approaches have been explored to minimise web stack interoperation.

This paper explores a pragmatic approach to reducing web stack interoperation, namely eliminating a tier/component. That is, we explore the implications of eliminating the Apache web server in a JAPyL web stack: Jupyter Notebook, Apache, Python, Linux, and replacing it with PHP libraries. We conduct a systematic study to investigate the implications for web stack performance, resource consumption, security, and programming effort.
 
\end{abstract}

\section{Introduction}

\begin{figure}[h!]

\includegraphics[width=\linewidth]{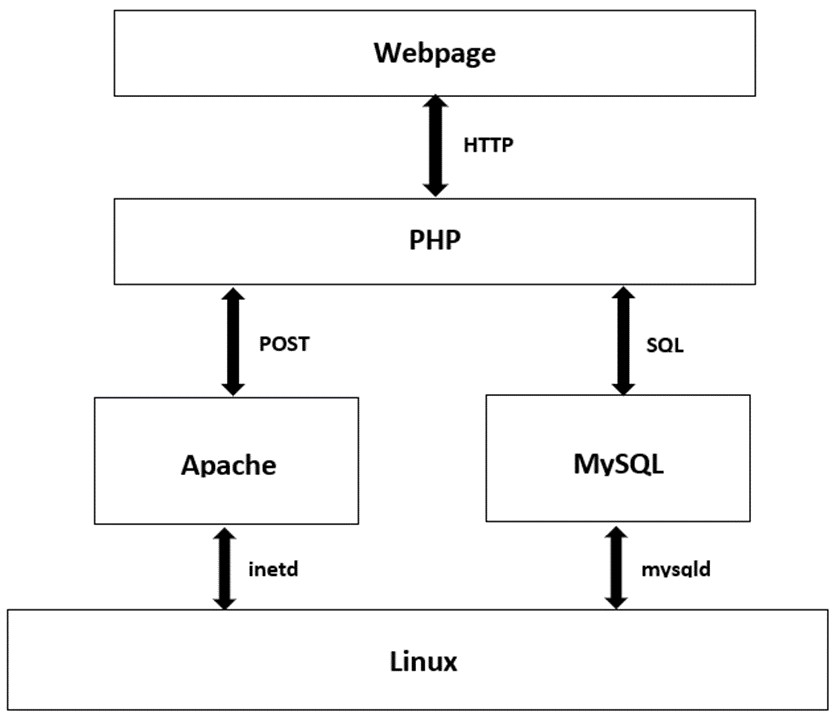}
\caption{4-Tier LAMP Architecture}
\label{fig:lamp}

\end{figure}

The architecture of a modern web application comprises a multi-tier stack of components.  The classic example is the 4-tier Linux, Apache, MySQL, and PHP (LAMP) stack~\cite{lee2003open}, cf.\ Figure~\ref{fig:lamp}. The components are written in different languages and interoperate using standard protocols like HTTP or SQL.

%Components, such as Apache, are independent, high performance, multi-threaded units or software with a high level of abstraction, designed to perform a specific function. On the other hand, the languages, such as PHP, can be seen as general purpose dynamic, scripting tools with a low level of abstraction necessary to deal with all the semantics, paradigms, libraries, modules, objects etc. This is needed in order to create the necessary interfaces to allow users to interact with the applications and components \cite{szyperski2002component}.

%The interoperation between components and languages within the web stack, as shown in Figure \ref{fig:lamp}, usually take two forms:

%\begin{enumerate}

%\item Component Interoperability - components designed to not only solve perform a specific function but co-operate with other components as part of an entire software system\\

%\item Language Interoperability - utilises all the necessary APIs, classes, objects, etc. necessary to facilitate smooth interoperation of the stack architecture. It focuses as well on how to cope and ensure cooperation of the components.

%\end{enumerate}

%Usually, the components interoperate through several methods such as dynamic on-load or on-demand linking or even through simple configuration files. On the other hand, languages such as PHP \& MySQL are made interoperable through several mechanisms such as objects, libraries, modules, etc.
A diverse set of components and languages raises a number of challenges. (1) Interoperation increases developer effort: the developer must be fluent in all of the languages, components and their interactions, i.e.~ be a full stack developer. (2) Interoperation can introduce security vulnerabilities like SQL injection. (3) Interoperation reduces performance as requests must be handled by multiple components, and data must be marshalled between them. (4) Interoperating multiple components consumes additional resources as memory and compute time is required for each component.

A range of approaches have been used to improve interoperation in web stacks. Some stacks focus on a single language, e.g.\ MEAN focuses on Javascript~\cite{dayley2014node}. Other stacks use a common VM to minimise the interoperation overheads between languages, e.g. the .NET framework uses the Common Language Runtime~\cite{gough2001compiling}. Sometimes web stack languages are combined, e.g. PyHyp combines Python and PHP~\cite{barrett2015fine}. The most radical approach is to combine all web stack languages into a single \emph{tierless} language, such as Links~\cite{cooper2006links} or Hop~\cite{serrano2006hop}.

%take a radical approach to overcoming some these challenges. They eliminate interoperation by providing  a single language that provides the functionality of all layers of the stack. 
%Tierless languages have not, however, been widely adopted.

%However, tierless languages have not been entirely successful in solving all associated interoperability challenges. There are cases, as shown in Appendix A, where these programming languages have compounded the problems. As a result, the use of different web stacks remain common.

We take a less radical but more pragmatic approach to simplifying web stacks. We investigate the implications of replacing a tier or component, with code libraries. That is, we explore the implications of eliminating the Apache web server in a Jupyter Notebook web stack called JAPyL (Jupyter, Apache, Python, Linux), replacing it with a PHP threaded library.

%The aim is to determine if reduced interoperation enables the construction of simpler, more performant, more efficient and more secure application stacks. As a result, this paper explores the following hypotheses.

This paper explores the following hypotheses for web stack tiers. (1) Does reducing the number of tiers improve web stack performance i.e.\ latency, throughput. (Section~\ref{sec:per})? (2) Does reducing the number of tiers reduce resource consumption, i.e.\ core utilization and memory usage (Section~\ref{sec:res})? (3) What are the security implications of eliminating a tier (Section~\ref{sec:security})? (4) Does reducing the number of tiers  make the construction of web applications easier, i.e.\ lines of code, cyclomatic complexity, size of interoperation code (Section~\ref{sec:prom})? %and specifically latency (time to respond to a request), and throughput (how many requests can be handled/s) (2) Does reducing the number of components or tiers reduce resource consumption?specifically core utilization, memory overhead? (3) What are the security implications of eliminating web stack components? (4) Does reducing the number (shrinking the depth?) of web stack components and tiers through language interoperability ease the construction of web applications? \pwtcomment{How do we measure: how many lines of code? how complex is code interoperation?}

As a basis for investigating these hypotheses we construct and compare two web stacks. The first, JAPyL, is a conventional 4-tier stack comprising a Jupyter notebook, an Apache web server, and a Python kernel, running on Linux (Section~\ref{sec:JAPyL}). The second, JPL, is a 3-tier stack that eliminates the Apache server and comprises a Jupyter notebook, PHP React (thread) library and kernel on Linux (Section~\ref{sec:JPL}). 

%\paragraph{Study Specifics}We compare the following aspects of our case study applications, to produce some surprising results. 
%\begin{itemize}
%    \item Performance i.e. latency, throughput. (Section~\ref{sec:per})\\
%    \item Resource Consumption i.e. Core Utilization, Memory Overhead(Section~\ref{sec:res})\\
%    \item Security Implications i.e. how to avoid attacks, declarative vs imperative programming, abstraction levels.
%    (Section~\ref{sec:security})\\
%    \item Programmability i.e. lines of code, cyclomatic complexity, size of interoperation  (Section~\ref{sec:prom})\\
%\end{itemize}

\section{Related Work}

\subsection{Interoperability Challenges}

Traditional web stacks like LAMP require the developer to interoperate a number of components, and this raises a number of  challenges.

\textbf{Increased Developer Effort.} The developer must effectively use multiple components and languages, and their interfaces
%in the stack requires the developer to either be well versed or at least fairly knowledgeable of the different programming paradigms, installation requirements and component functionality
i.e. be a full stack developer~\cite{northwood2018full}.%In addition, the developer may have to find innovative ways such as through APIs or unstructured techniques to facilitate the interoperation especially if compatibility is lacking \cite{northwood2018full}.\\ %One classic example of this is the Full Stack Developer who can code at every tier of an application stack. This is essentially a person who is both knowledgeable in the inner workings of databases, web servers, operating systems, app code development, etc. and has the necessary skillset to  successfully interoperate them\cite{northwood2018full}.\\

\textbf{Semantic Friction}, or computational impedance mismatch,  arises when the developer must interoperate different programming language paradigms. For example, many web stacks interoperate an imperative programming language with a relational database language, e.g. PHP \& SQL ~\cite{ireland2009classification}. %This may ultimately result in unnatural programming structures or more programming effort being needed \cite{ireland2009classification}.\\

\textbf{Security Vulnerabilities.} Interoperating multiple components increases the attack surface of the application, e.g. in LAMP, interoperating  Apache with PHP can leave the stack vulnerable to Cross Site Scripting (XSS) if security headers are not implemented or web forms are not properly sanitized~\cite{endler2002evolution}.

%\item Security Vulnerabilities - in the Web 2.0 era, cyberattacks such as Cross Site Sciprting (XSS) attacks have become all too common. This is due in part to the use of certain components and languages. For example, with Apache \& PHP, there are two attack vectors which attackers usually examine to determine if the web stack is vulnerable to XSS. The first is to examine if the security header HTTP X-XSS-Protection has not been enabled in Apache as this is used by the client browser when loading a web application to detect and prevent reflected XSS. The other is to insert a malicious script into a PHP webform and examine if it will be executed especially if the form is not properly sanitized.  \cite{endler2002evolution}.\\

\textbf{Performance Overheads} can be introduced by interoperation, e.g. data must be marshalled between components~\cite{li2013jvm}, and memory and compute resources must be available for each component.

%\item Performance Overhead - interoperating languages or components usually incurs a performance overhead. For example, inter-process communication is typically around five to six times slower than a function call as data must be marshalled for communication between languages or components. The reason is that time is either needed for the different languages to communicate with one another or to translate messages from one language to the next \cite{li2013jvm}.

\subsection{Reducing Webstack Interoperation}

Some approaches to reducing interoperation in  modern multi-tier web stacks are as follows.

Interoperation is reduced by \textbf{consistently using a single language} as far as possible  throughout the stack. For example, the MEAN web stack uses JavaScript as the primary programming language and composes MongoDB, ExpressJS, AngularJS \& NodeJS~\cite{dayley2014node}. MEAN typically outperforms LAMP by 2.5x or more due to the computational inefficiencies of PHP, and the inefficiency of Apache in handling I/O operations ~\cite{chaniotis2015node}.

A \textbf{Common Language Runtime} or virtual machine can ease interoperation. For example, the .NET Framework typically composes WinForms, ADO.NET, Framework Class Library and the Common Language Runtime (CLR)~\cite{gough2001compiling}. Having the CLR as a  common runtime facilitates the interoperation of a number of Microsoft technologies like ASP.NET, VB.NET, C\#, F\#, etc. However .NET is proprietary and only available on the Windows Operating System.

Interoperating a wider range of languages on a common VM is also possible, e.g.~the Truffle framework interoperates JavaScript, Ruby, Python, JAVA, Scala, C++, etc. It does so using the GraalVM compiler and  an abstract syntax tree interpreter.  The platform is, however, experimental and has some performance overheads, e.g.  the Truffle VM does not marshal objects at the language boundary but rather directly passes them from one language to the next ~\cite{grimmer2018cross}.

\textbf{Combining web stack languages.} Interoperation in our JPL webstack could be reduced by using PyHyp~\cite{barrett2015fine}, a language that combines the Python used in Jupyter and PHP. PyHyp is, however, relatively immature and JPL uses Python and PHP as established technologies.

\textbf{Tierless Languages.} The most radical approach is to combine \emph{all} of the webstack languages into a single tierless language, as exemplified by 
 languages like  Links~\cite{cooper2006links} or Hop~\cite{serrano2006hop}. Here developers write both client and server as a single program in a single language. The client and server code is simultaneously checked by the compiler, and compiled to the required component languages. For example, Links compiles to HTML \& JavaScript for the client side and to SQL on the server-side to interact with the database system. While conceptually appealing, tierless languages face some challenges in practice. For example, Appendix A~\cite{ramsingh20} shows an instance where Links requires additional developer effort compared to PHP and MySQL. Tierless languages have had limited uptake to date, and tiered architectures remain commonplace.

\subsection{Hosting Jupyter Notebooks}

The Jupyter Platform has gained in popularity especially in the realm of data science because it allows users to process, analyse \& manipulate data; and create analytical and statistical models with just a few lines of code that can be tested straight from a browser user interface. In addition, Jupyter Notebooks allows for the interoperation of different languages, such as the popular R-Python combination used by data scientists, through custom techniques such as subkernels and magics.

\begin{figure}[h!]

\includegraphics[width=\linewidth]{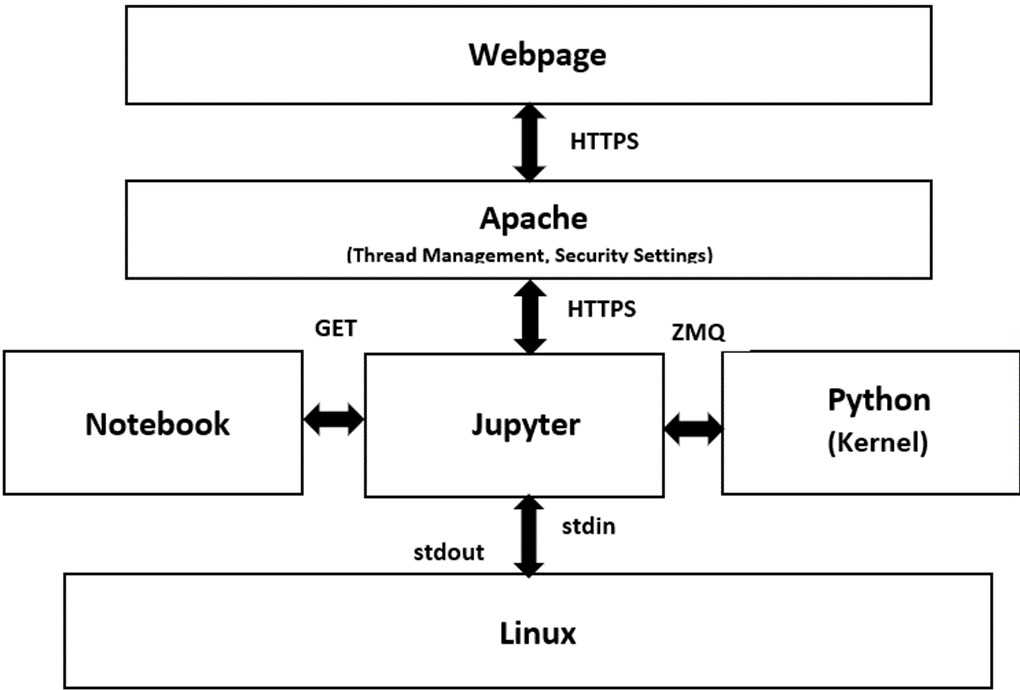}
\caption{4-Tier JAPyL Architecture}
\label{fig:japyl}

\end{figure}

Generally, the standard way to make Jupyter Notebooks web accessible is to use the Jupyter, Apache, Python, Linux (JAPyL) web stack (Figure~\ref{fig:japyl}), and embed the Notebook into a webpage or site with an Apache Web Server as a Reverse Proxy~\cite{ristic2005apache}. Here, Apache is usually interposed between the client and the Jupyter Server, taking requests from clients and forwarding them to Jupyter. This is done in order to create an extra layer of security to protect the Jupyter Server~\cite{milligan2017interactive}. Appendix B~\cite{ramsingh20} provides an example of the configurations a developer has to implement in both Apache and the Jupyter Server to allow for communication between the two components.

%From a security perspective, this architecture is very important by design because Jupyter Notebooks pose several security risks if not properly secured online. For example, if Jupyter is not configured properly and an attacker is able to get access to the Notebook code environment, it becomes easy for the attacker to execute malicious commands. The reason for this is that Jupyter Notebooks do not sanitize or check the code being executed\cite{milligan2017interactive}. 

%\section{Alternative Technology Stacks}

\section{Case Studies}

\label{sec:studies}

%\subsection{Our Approach}

We explore the implications of eliminating a tier in a web stack, focusing on how the interoperation changes. Specifically we compare two Jupyter notebook stacks: JAPyL that composes Jupyter, Apache, Python and Linux in 4 tiers (Figure~\ref{fig:japyl}); and JPL that replaces Apache with a PHP thread library and composes Jupyter, PHP and Linux (Figure~\ref{fig:jpl}) in 3 tiers.

\subsection{JAPyL}
\label{sec:JAPyL}
JAPyL is a conventional 4-tier architecture composing Jupyter, Apache, Python and Linux (Figure~\ref{fig:japyl}).

\subsubsection{Security Configurations}

\begin{figure}[h!]

\includegraphics[width=\linewidth]{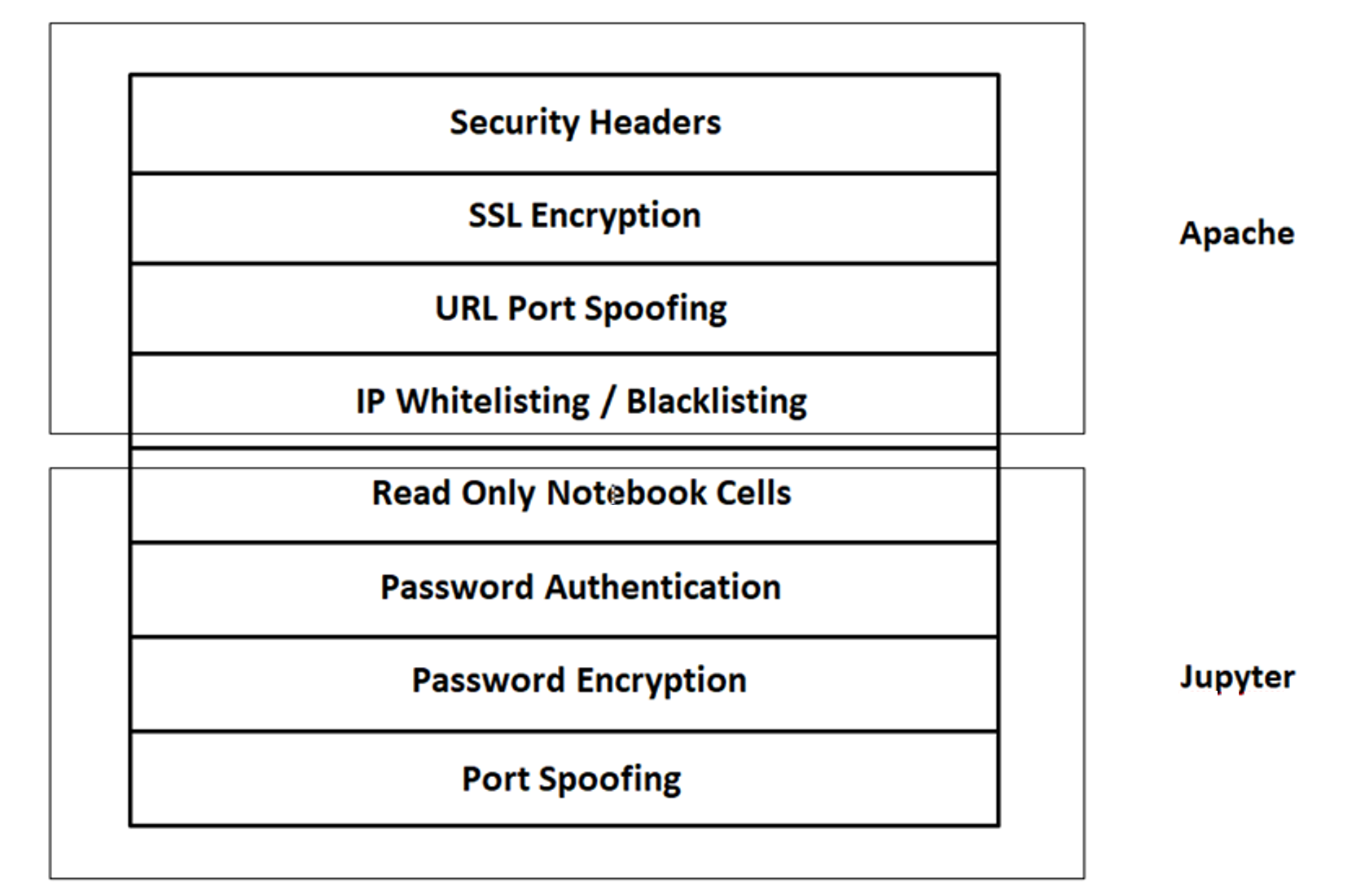}
\caption{JAPyL Defense-in-Depth Security Model}
\label{fig:dip}

\end{figure}

In addition to the Reverse Proxy JAPyL also utilises a Defense-in-Depth multi-layered security approach (Figure~\ref{fig:dip}). That is, various security mechanisms are deployed throughout the stack. The intention is that if an attacker targets the Jupyter Notebook online and is able to penetrate one layer, another layer may thwart the attack.

The JAPyL Defense-in-Depth model comprises: (1) 
Security Headers (2) SSL Encryption (3) URL Port Spoofing (4) IP Whitelisting / Blacklisting (5) Read Only Notebook Cells (6) Password Authentication (7) Password Encryption (8) Port Spoofing.

\subsubsection{Declarative Configuration and Security}

The security mechanisms are typically declaratively specified in either the Jupyter Server or Apache~\cite{lloyd1994practical}. Often security features are implemented as runtime parameters. Appendix C~\cite{ramsingh20} provides examples of some of the declarative security mechanisms used in JAPyL.

It is commonly believed that such declarative specifications in some domain-specific language such as HTML, XML, or Apache Configs makes things easier for the developer by  raising the level of abstraction. Nonetheless the developer must be fluent in a range of languages, and for JAPyL these are  (1) Jupyter Configurations (2) Jupyter Notebook Interoperability Techniques i.e.\ Kernel, Subkernel \& Magic Implementations (3) Several different languages including JSON, Apache Configs, etc ~\cite{milligan2017interactive}.

\subsection{JPL}
\label{sec:JPL}
\begin{figure}[h!]

\includegraphics[width=\linewidth]{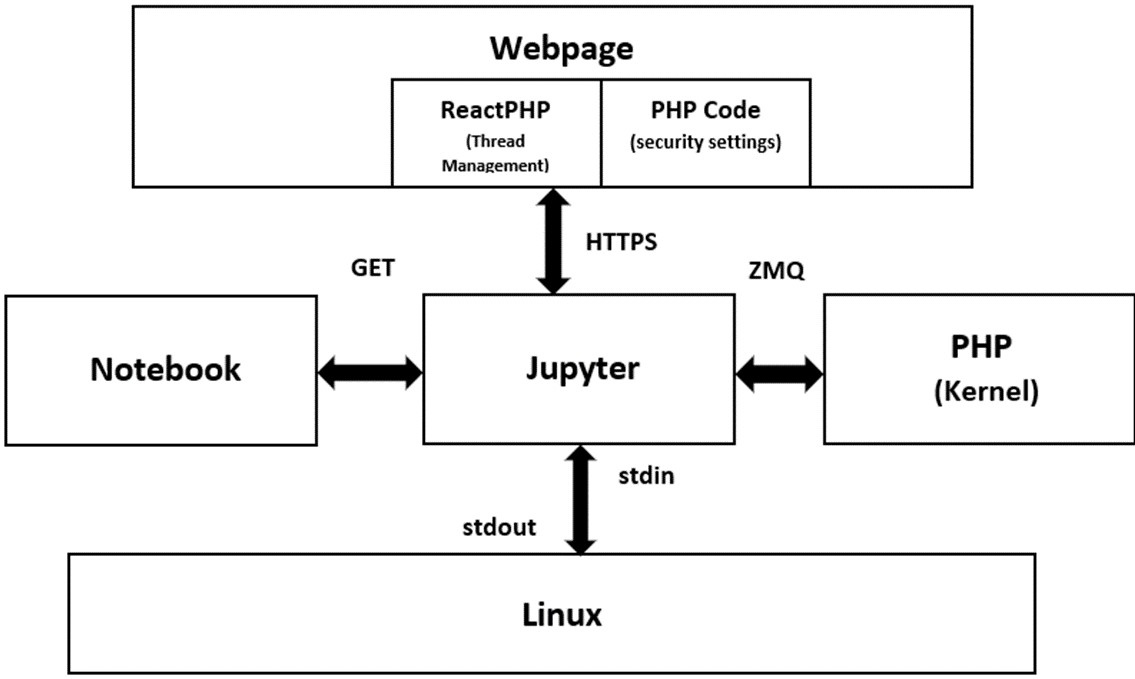}
\caption{3-Tier JPL Architecture}
\label{fig:jpl}

\end{figure}

JPL (Jupyter, PHP, Linux) is a 3-tier architecture that carefully replicates the web service and security mechanisms of JAPyL (Figure ~\ref{fig:jpl}). The Apache tier is replaced by the ReactPHP thread library, and some hand-coded PHP security code. 

\subsubsection{Imperative Configuration and Security}

In contrast to JAPyL's declarative specification of configuration and security much of the implementation is imperative. 
That is, only PHP was written to perform the necessary configurations as demonstrated by the examples in Appendix E~\cite{ramsingh20}.

Using an imperative paradigm means that the level of abstraction has decreased compared with JAPyL. However the developer has the expressiveness to implement the necessary functionalities, and can potentially implement features not supported by the JAPyL DSLs~\cite{freeman1990kaleidoscope}. 
%In other words, the programmer implements code and configurations telling the web application what to do and how to do it. 
%Thus the level of abstraction for configuration and functionality implementation is decreased in JPL compared with JAPyL.

\subsubsection{Why PHP?}

The rationale for using PHP to eliminate a web component is twofold. Not only is it one of the most popular and mature web programming technologies, but also offers a range of technical benefits~\cite{lerdorf2002programming}, as follows.
% Efficiency - uses its own memory space and thus decreases the loading time and workload from the server.\\
PHP provides simple parsing and marshalling, e.g. parsing JSON and XML with a single line of code.
PHP supports multiple major databases including MySQL, dBase, IBM DB2, InterBase, FrontBase, ODBC, PostgreSQL, SQLite, etc. 
PHP is a mature technology, and supported by a range of frameworks like CakePHP, CodeIgniter, Zend, Larvarel that not only make development faster but also provide flexible coding styles and interfaces for programmers. 

For our interoperability study, the most significant benefit of PHP is that it follows the familiar object oriented paradigm. Hence developers can, for example, create custom classes, and interoperate relatively smoothly with other object oriented languages in the stack like Java and, crucially for our study,  Python. 

Languages that are very similar  interoperate with less semantic friction. Python and PHP are indeed similar, and have even been combined into PyHyp~\cite{barrett2015fine}. That is both languages support class creation, encapsulation, functions, immutable data, inheritance, object creation, polymorphism, and shared state and shared memory.

%Furthermore, PHP has built in functions and features which makes its compatible to interoperate with other languages and components especially with Python, as shown in Table \ref{table:interop},  which is one of the core languages used in Jupyter. As Barrett et al. have shown from their work in combining PHP with Python to form a new language called PyHyp, languages that are very similar in structure or design have better interoperation performance and less semantic friction \cite{barrett2015fine}.

%\begin{table}[ht]
%\caption{PHP \& Python Similarities} % title of Table
%\centering % used for centering table
%\begin{tabular}{c c c} % centered columns (3 columns)
%\hline\hline %inserts double horizontal lines
%Feature & PHP & Python \\ [0.5ex] % inserts table
%heading
%\hline % inserts single horizontal line
%Class Creation & Yes & Yes \\ % inserting body of the table
%Encapsulation & Yes & Yes \\
%Functions & Yes & Yes \\
%Immutable Data & Yes & Yes \\
%Inheritance & Yes & Yes \\
%Object Creation & Yes & Yes \\
%Polymorphism & Yes & Yes \\
%Shared State & Yes & Yes \\
%Shared Memory & Yes & Yes \\
%\hline %inserts single line
%\end{tabular}
%\label{table:interop} % is used to refer this table in the text
%\end{table}

\section{Evaluation}

\subsection{Experiment Design}

We evaluate the performance and programmability of JPL by comparing it to JAPyL stacks on two platforms: Docker \& native Raspberry Pi 3. While Jupyter Stacks are mostly deployed in virtualized environments like Docker~\cite{zonca2018deploying}, it is far easier to obtain accurate core and memory resource measurements on the Raspberry Pi. Where possible, the components used in the JAPyL and JPL stacks are identical, and as follows: Docker 18.09, Raspbian Stretch, Jupyter Server 5.7.8, Ubuntu 16.04. JAPyL uses Apache Server 2.4.34, and JPL ReactPHP 0.8.4. Experiments for the Docker platforms is conducted on an Intel Core i3 system, Windows 10 Operating System with 2.23GHz and 8GB of RAM.

Two existing Jupyter Notebooks not written by the authors are downloaded from GitHub. They are selected to be well designed~\cite{rule2018exploration}, i.e. to have a notebook title \& introduction, descriptions of the model parameters, and of the data parameters, and to import packages. The notebooks are simply loaded by the JAPyL and JPL stacks.

To minimise variability, the reported results are the median of three consecutive benchmark executions.

\subsection{Performance: Latency and Throughput}
\label{sec:per}

Given the reduced number of tiers in JPL compared with JAPyL and that PHP is similar to Python, it is reasonable to expect JPL to outperform JAPyL.

\begin{figure}[h!]

\includegraphics[width=\linewidth]{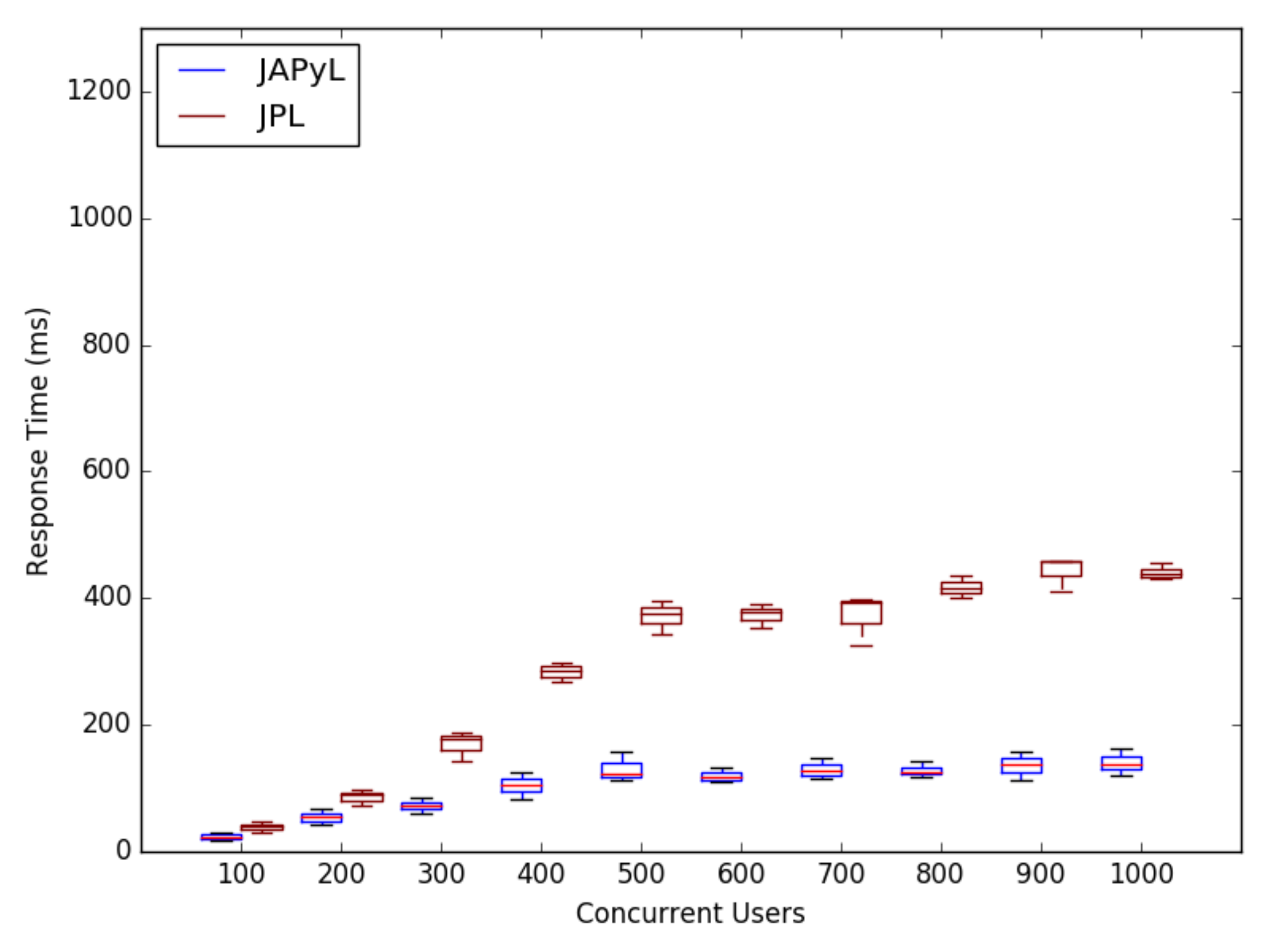}
\caption{JAPyL vs JPL Latencies (Docker)} %\pwt{On what platform?}
\label{fig:latencydock}

\end{figure}

\begin{figure}[h!]

\includegraphics[width=\linewidth]{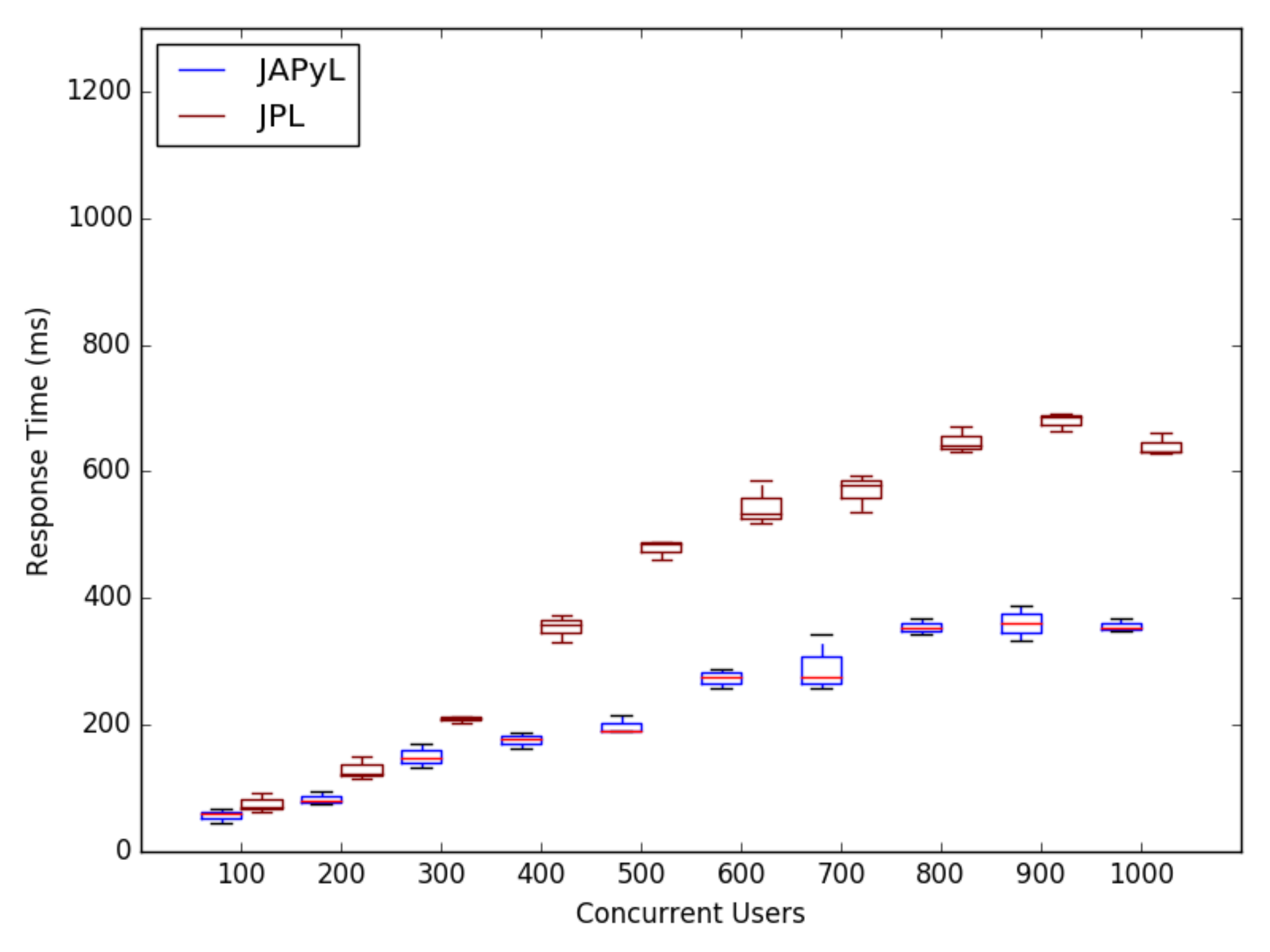}
\caption{JAPyL vs JPL Latencies (RaspberryPi)} %\pwt{On what platform?}
\label{fig:latencyrasp}

\end{figure}

Figure ~\ref{fig:latencydock} shows the request latencies of JAPyL and JPL as the number of concurrent connections varies from 50 to 1000 on all platforms.  Contrary to our expectations it shows that JPL latency is  two or three times greater than JAPyL. The results are almost similar for the RaspberryPi as shown in Figure~\ref{fig:latencyrasp}.

\begin{figure}[h!]

\includegraphics[width=\linewidth]{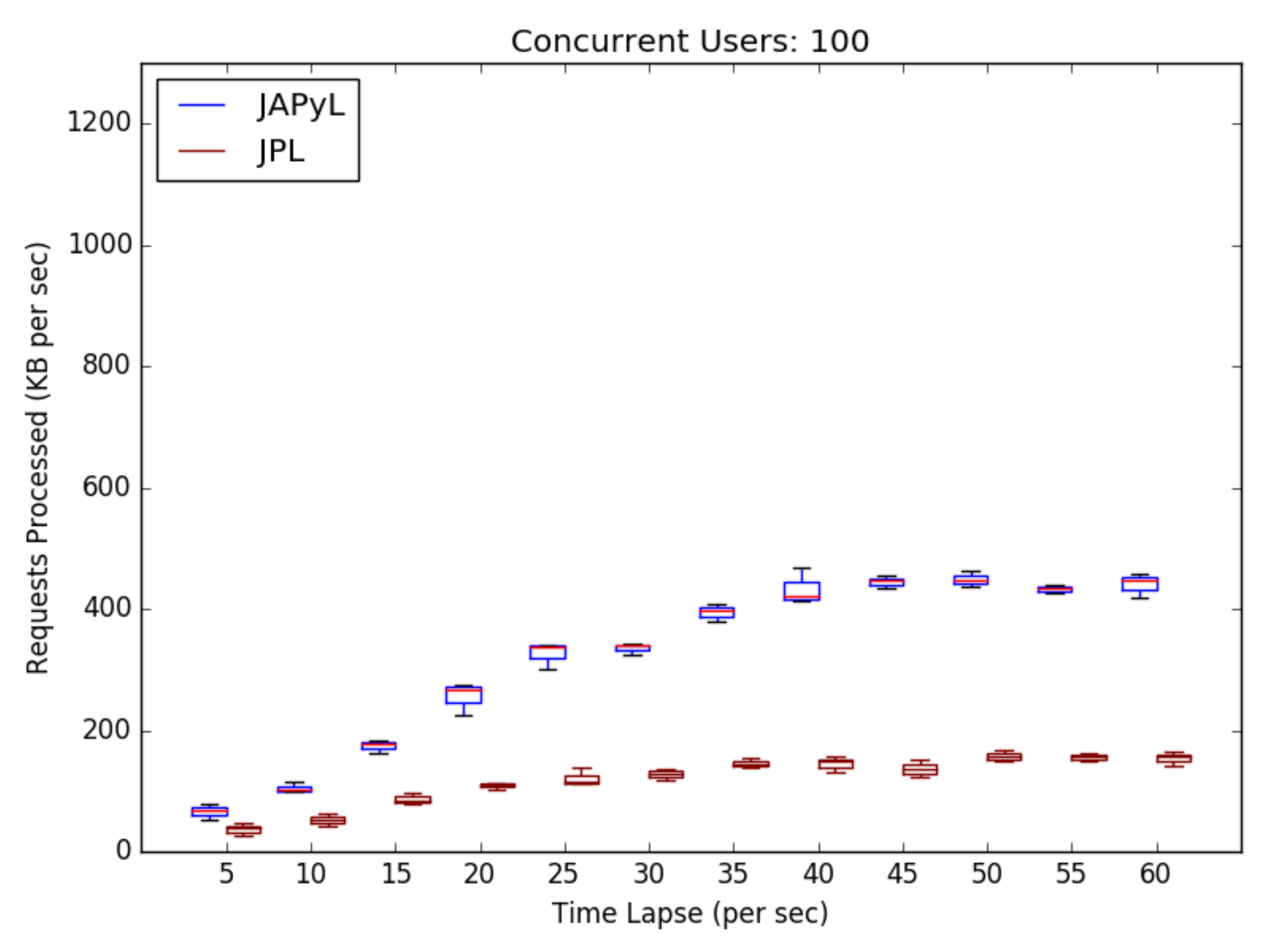}
\caption{JAPyL vs JPL Throughput (Docker)} %\pwt{On what platform?}
\label{fig:throughputdock}

\end{figure}

Figure ~\ref{fig:throughputdock} shows the request throughput of JAPyL and JPL as the number of connections varies. It also confounds our expectation by revealing that JPL throughput is two orders of magnitude lower than for JAPyL. We attribute this to relatively poor thread management in the PHP React library~ ~\cite{lerdorf2002programming}, compared with Apache that is designed for high degrees of parallelism and to effectively utilise multicore systems~\cite{lee2003open}. \\

\begin{figure}[h!]

\includegraphics[width=\linewidth]{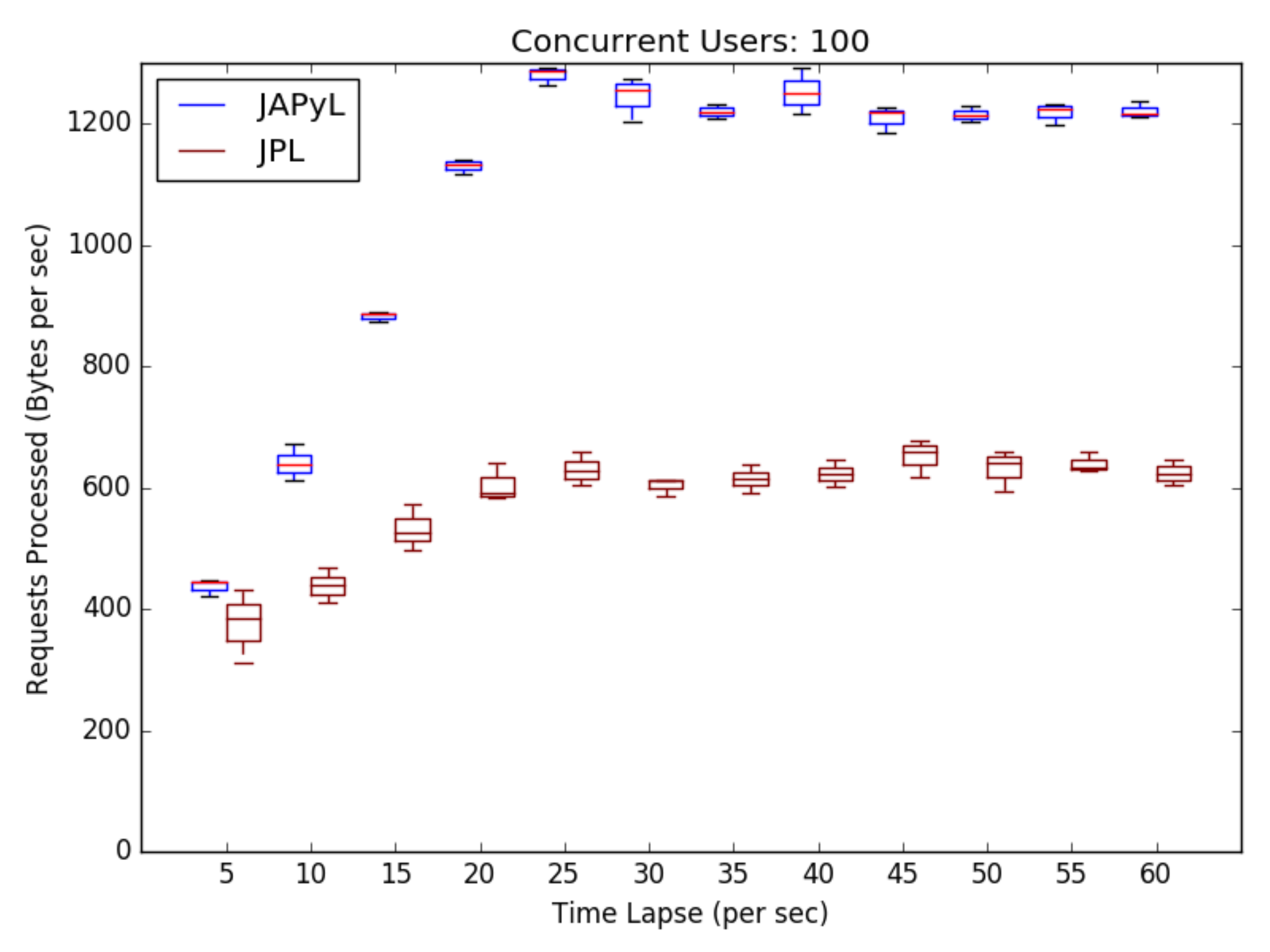}
\caption{JAPyL vs JPL Throughput (RaspberryPi)} %\pwt{On what platform?}
\label{fig:throughputrasp}

\end{figure}

%The reason for this slow performance could be that despite the advances made by PHP in terms of concurrency, the language itself is still naturally not quite oriented towards multi-threading. In other words, it still cannot spawn more than a few processes at a time as previous versions of PHP were single-threaded. \cite{lerdorf2002programming}.

%Compare this to Apache which has several sub-processes running in parallel one for each request, up to a couple of dozens or hundreds, depending on the configuration \cite{lee2003open}. Put simply, Apache is built for concurrency, parallelism and to utilise multicore systems.

%The results are very similar for the VirtualBox and RaspberryPi platforms (Appendix F~\cite{ramsingh20}).%(Figures ~\ref{fig:latencyvm}, ~\ref{fig:latencyrasp}, ~\ref{fig:throughputvm}, ~\ref{fig:throughputrasp} 

\subsection{Resource Usage: Core Utilization, Memory Overhead}
\label{sec:res}

%\pwtcomment{Should this be core rather than CPU?}

As the JPL stack eliminates Apache and runs fewer components it is reasonable to expect JPL to consume less resource than JAPyL.

\begin{figure}[h!]

\includegraphics[width=\linewidth]{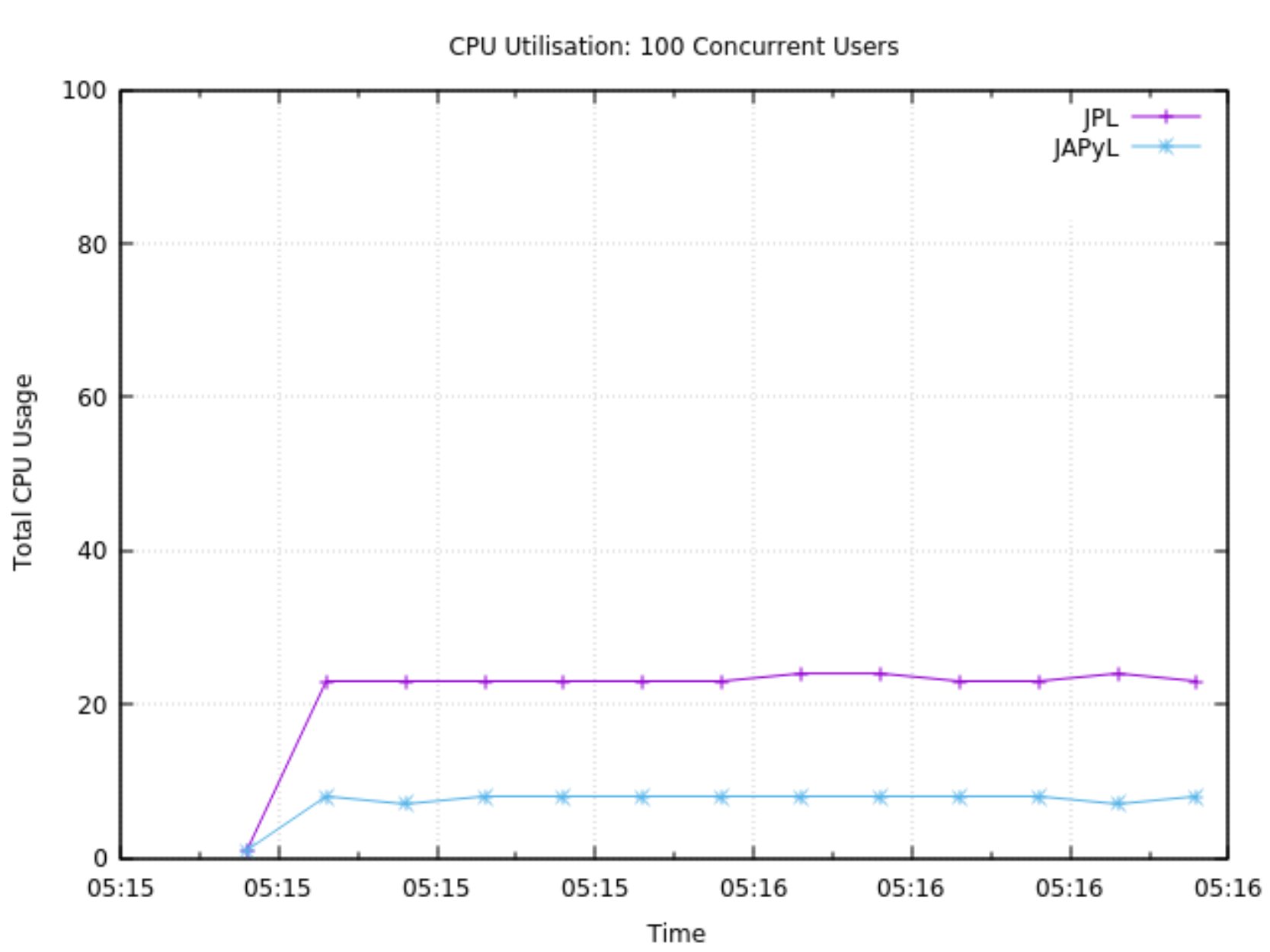}
\caption{JAPyL vs JPL CPU Resource Utilisation (RaspberryPi)} %\pwt{On what platform?}
\label{fig:cpu}

\end{figure}

Figure~\ref{fig:cpu} shows the CPU utilisation of JAPyL and JPL as they process https requests with 100 concurrent connections on the Raspberry Pi platform. Again, contrary to expectation JPL utilisation is typically 20\% greater than JAPyL.

A possible explanation for this is memory leaks. This refers to a long running PHP request where the amount of memory utilised will slowly increase over time. There is some memory in PHP which just cannot be freed up on a regular basis due to its reliance on reference counting to manage memory~\cite{lerdorf2002programming}. 

%\begin{figure}[h!]
%\includegraphics[width=\linewidth]{cpu_100.png}
%\includegraphics[width=\linewidth]{cpu_500.png}
%\caption{Core Utilization (Raspberry Pi3)} %\pwt{On what platform?}}
%\label{fig:core}

%\end{figure}

%\pwtcomment{I have serious doubts about Figure 11. Shall we drop it and report  only the memory residencies below?}%

\paragraph{Memory Usage}

Table ~\ref{table:size} shows the memory residencies of the JAPyL and JPL on the Docker Platform while being initialized and hosting the relevant notebooks. With fewer components in JPL, overall size might be expected to be smaller, but it is 135MB greater. Table ~\ref{table:object} reveals why:  PHP is much larger than Apache. 

\begin{table}[ht]
\caption{Memory Residencies} % title of Table
\centering % used for centering table
\begin{tabular}{c c c c c} % centered columns (4 columns)
\hline\hline %inserts double horizontal lines
Stack & Size(MB) \\ [0.5ex] % inserts table
%heading
\hline % inserts single horizontal line
JAPyL & 645\\ % inserting body of the table
JPL & 860\\
\hline %inserts single line
\end{tabular}
\label{table:size} % is used to refer this table in the text
\end{table}

\begin{table}[ht]
\caption{Component Size} % title of Table
\centering % used for centering table
\begin{tabular}{c c c c c} % centered columns (4 columns)
\hline\hline %inserts double horizontal lines
Object & Size(MB) \\ [0.5ex] % inserts table
%heading
\hline % inserts single horizontal line
Apache & 165\\ % inserting body of the table
PHP & 390\\
\hline %inserts single line
\end{tabular}
\label{table:object} % is used to refer this table in the text
\end{table}

%Memory leaks refer to a long running PHP request where the amount of memory utilised will slowly increase over time. There is some memory in PHP which just cannot be freed up on a regular basis due to its reliance on reference counting to manage memory. Garbage collection is just utilised to pick up the pieces that the reference counter misses. The most common source of memory leaks are cyclic references and global variables\cite{lerdorf2002programming}. As a result, the longer a PHP request takes, the more memory it leaks which results in further memory overhead.

\subsection{Security}
\label{sec:security}

Table ~\ref{table:security} shows that the JAPyL Defense-in-Depth security model can be replicated in JPL, but requires key functions to be hand coded. In some cases, such as the security header feature, JPL was able to surpass JAPyL as the implementation could be automated with just a few lines of code as demonstrated in Appendix E~\cite{ramsingh20}. That is, hand coding enables greater security than relying on Apache's declarative security options.

\begin{table}[ht]
\caption{JAPyL vs JPL Security Implementations} % title of Table
\centering % used for centering table
\begin{tabular}{c c c c c} % centered columns (4 columns)
\hline\hline %inserts double horizontal lines
%heading
\hline % inserts single horizontal line
Read Only Cells & JSON Coding & JSON Coding \\ % inserting body of the table
Password Encryption & Command Line & Command Line \\
Port Spoofing & Hand Coded & Hand Coded \\
SSL Encryption & Apache Configs & Hand Coded \\
IP Whitelisting & Hand Coded & Hand Coded \\
IP Blacklisting & Hand Coded & Hand Coded \\
Security Headers & Hand Coded & Automated\\
\hline %inserts single line
\end{tabular}
\label{table:security} % is used to refer this table in the text
\end{table}

%In this context, the automation of security headers through the use of the imperative programming paradigm in JPL reveals that low levels of abstraction are actually a good thing. It can allow the developer the control, freedom and expressiveness to implement functionalities or interoperate components \& languages in innovative ways that are actually better than declarative programming.

\subsection{Programmability}
\label{sec:prom}

\subsubsection{Code Size}

Code size is widely recognised as a measure of developer effort and of maintainability. Table~\ref{table:loc} enumerates the lines of code required to implement the functionalities of the JAPyL and JPL stacks. Implementing the JPL web stack requires 267 fewer lines of code, or 42\% less code.  This is to be expected because there are fewer languages and paradigms to be implemented as shown in Table~\ref{table:langparadigm}. JAPyL utilises 3 programming languages compared to 2 in JPL.

\begin{table}[ht]
\caption{Lines of Code (LOC)} % title of Table
\centering % used for centering table
\begin{tabular}{c c c c c} % centered columns (4 columns)
\hline\hline %inserts double horizontal lines
Functionality & JAPyL & JPL \\ [0.5ex] % inserts table
%heading
\hline % inserts single horizontal line
Embed Notebook & 10 & 36 \\ % inserting body of the table
Host Webpage & 29 & 26\\
Reverse Proxy & 33 & 23\\
Security Configs & 21 & 31\\
Language Processing & 527 & 63\\
Messaging & 11 & 85\\
\hline
Total & 631 & 364\\
\hline %inserts single line
\end{tabular}
\label{table:loc} % is used to refer this table in the text
\end{table}

%\begin{table}[ht]
%\caption{Programming Language Implementation} % title of Table
%\centering % used for centering table
%\begin{tabular}{c c c c c} % centered columns (4 columns)
%\hline\hline %inserts double horizontal lines
%Functionality & JAPyL & JPL \\ [0.5ex] % inserts table
%heading
%\hline % inserts single horizontal line
%Embed Notebook & HTML & PHP \\ % inserting body of the table
%Host Webpage & Apache Configs & PHP\\
%Reverse Proxy & Apache Configs & PHP\\
%Security Settings & Apache Configs & PHP\\
%Language Processing & Python & PHP\\
%Messaging & JSON & PHP\\
%\hline
%Total & 4 & 1\\
%\hline %inserts single line
%\end{tabular}
%\label{table:lang} % is used to refer this table in the text
%\end{table}

\begin{table*}
	\centering
	\caption{Implementation Languages and Paradigm Comparison.}%
	\label{table:langparadigm}
	%\mlcomment{If we were really desparate for space we could even merge in table 3}}%
	\begin{tabular}{lllll}
		\hline
        \hline
		                     & \multicolumn{2}{c}{Languages}  & \multicolumn{2}{c}{Paradigms}\\
		\hline
		Functionality        & JAPyL   & JPL & JAPyL              & JPL\\
		\hline
		Embed Notebook        & HTML       &  HTML      & Declarative             & Declarative \\
		Host Webpage        & Apache Configs    & PHP      & Declarative & Imperative \\
		Reverse Proxy        & Apache Configs    & PHP      & Declarative & Imperative\\
		Security Settings & Apache Configs & PHP & Declarative  & Imperative\\
		Language Processing & Python & PHP & Object-Oriented  & Object-Oriented\\
		Messaging   & JSON       & PHP      & Data Serialization              & Imperative\\
		\hline
		Total                & 4           & 2          & 3                       & 2\\
		\hline
	\end{tabular}
\end{table*}

\subsubsection{Code Complexity}

%With fewer tiers the JPL developer writes less code, but how complex is the code? This is dependent on how many paradigms the programmer must use and how many control flows must be managed.

\paragraph{Language Implementation} Code complexity is dependent on how many paradigms the programmer must use and how many control flows must be managed~\cite{casti1986system}. Our expectation is that since JPL uses fewer programming languages and paradigms than JAPyL in implementation, the code complexity will be less.

%{The Complexity} of the code is measured using Cyclomatic Complexity~\cite{McCabe} which has the number of control paths as a key metric
However, our experiments reveal that the structural code complexity needed to interoperate the components and PHP language in JPL is slightly higher when compared to JAPyL based on Cyclomatic Complexity. Table~\ref{table:complex} shows that JAPyL has a rating of 19 when compared to 21 in JPL. This means that the code in JPL has more control flows that have to be managed.

\begin{table}[ht]
\caption{Cyclomatic Complexity (cc)} % title of Table
\centering % used for centering table
\begin{tabular}{c c c c c} % centered columns (4 columns)
\hline\hline %inserts double horizontal lines
Functionality & JAPyL & JPL \\ [0.5ex] % inserts table
%heading
\hline % inserts single horizontal line
Embed Notebook & 1 & 1 \\ % inserting body of the table
Host Webpage & 1 & 1\\
Reverse Proxy & 1 & 1\\
Security Configs & 1 & 5\\
Language Processing & 13 & 1\\
Messaging & 2 & 12\\
\hline
Total & 19 & 21\\
\hline %inserts single line
\end{tabular}
\label{table:complex} % is used to refer this table in the text
\end{table}

This is crucial because despite having fewer lines of codes, a higher complexity number means that the programmer may have to deal with more control paths in the JPL code which could lead to more unexpected results and defects such as poor performance and higher resource consumption due to interoperation. Casti et al.\ state that just because a system or application has fewer components or layers does not usually make make it simpler or less complex. You still have to take into account the processes and behaviour interactions which may be impacted as well as the compatibility factor between system components and languages~\cite{casti1986system}.

\section{Conclusion}

\paragraph{Summary} We have explored whether reducing the number of tiers/components eases the construction of web applications by reducing interoperability. We did so by systematically comparing the 4-tier JAPyL and 3-tier JPL web stacks  (Section~\ref{sec:studies}). The key findings from our case study are as follows.

\textbf{Performance.} Eliminating the Apache component, and associated interoperation, in JPL and replacing it with a PHP threaded library increases latency (Figure~\ref{fig:latencydock}) and reduces throughput (Figure~\ref{fig:throughputdock}). We believe this reflects that the Apache thread management is far superior to that provided by the PHP React library (Section~\ref{sec:per}).

\textbf{Resource Consumption} Despite replacing  Apache,  JPL consumes more resources. JPL uses 30\%  more core cycles, and  (Figure~\ref{fig:cpu}), and we believe that this is due to Apache being better at thread management. JPL uses 33\% more memory (Table~\ref{table:size}) as PHP is much larger (390Mb) than Apache (165Mb) (Section~\ref{sec:res}).

\textbf{Security} provided by JPL and JAPyL is almost identical, although hand-written PHP in JPL can provide additional capabilities, e.g. automating the handling of security headers. This is a benefit of the lower-level imperative security coding  (Section~\ref{sec:security}).

\textbf{Programmability} Eliminating the Apache tier/component results in the following. There is a reduction in number of programming languages and paradigms utilised (Table~\ref{table:langparadigm}) and a smaller code size (Table~\ref{table:loc}) when compared to JAPyL. This means less developer effort for stack implementation and reduced chances of semantic friction.

%\textbf{Programmability} Despite eliminating the Apache component, and associated interoperation, the developer effort in the two stacks is very similar, as measured by lines of code and cyclomatic complexity. The primary reason is that there is more hand written code in JPL, e.g.~for security (Section~\ref{sec:prom}).

\textbf{Reflection.} We anticipated that eliminating a tier/ component would simplify the stack. By reducing interoperation we hoped to improve performance and security, and to reduce resource consumption and programming effort. Our JAPyL/JPL study confounded these expectations. It is possible that substituting Apache with PHP was a poor decision, and that using  a language with better memory management and support for multi-threading, like Erlang or Go, would meet these expectations. 

\textbf{Future work} will further investigate replacing monolithic stack components with programming language technologies for a variety of stacks, domains, and programming languages. We intend that tier elimination has minimal impact on an application stack's function, but seek to explore potential non-functional benefits. To this end we are currently comparing a conventional, and a tierless Clean iTask/mTask,  implementation of a smart campus IoT stack.

\bibliographystyle{abbrv}
\bibliography{refs}

\begin{thebibliography}{10}

\bibitem{ramsingh20}
R.~Adrian, S.~Jeremy, and P.~Trinder.
\newblock Online appendix: Do fewer tiers mean fewer tears?, 2020.
\newblock \url{http://www.dcs.gla.ac.uk/~ramsad/papers/FewerTiersAppendix.pdf}.

\bibitem{barrett2015fine}
E.~Barrett, C.~F. Bolz, L.~Diekmann, and L.~Tratt.
\newblock Fine-grained language composition: A case study.
\newblock {\em arXiv preprint arXiv:1503.08623}, 2015.

\bibitem{casti1986system}
J.~L. Casti.
\newblock On system complexity: Identification, measurement, and management.
\newblock In {\em Complexity, language, and life: Mathematical approaches},
  pages 146--173. Springer, 1986.

\bibitem{chaniotis2015node}
I.~K. Chaniotis, K.-I.~D. Kyriakou, and N.~D. Tselikas.
\newblock Is node. js a viable option for building modern web applications? a
  performance evaluation study.
\newblock {\em Computing}, 97(10):1023--1044, 2015.

\bibitem{cooper2006links}
E.~Cooper, S.~Lindley, P.~Wadler, and J.~Yallop.
\newblock Links: Web programming without tiers.
\newblock In {\em International Symposium on Formal Methods for Components and
  Objects}, pages 266--296. Springer, 2006.

\bibitem{dayley2014node}
B.~Dayley.
\newblock {\em Node.js, MongoDB, and AngularJS web development}.
\newblock Addison-Wesley Professional, 2014.

\bibitem{endler2002evolution}
D.~Endler.
\newblock The evolution of cross site scripting attacks.
\newblock Technical report, Technical report, iDEFENSE Labs, 2002.

\bibitem{freeman1990kaleidoscope}
B.~N. Freeman-Benson.
\newblock Kaleidoscope: mixing objects, constraints, and imperative
  programming.
\newblock In {\em ACM SIGPLAN Notices}, volume~25, pages 77--88. ACM, 1990.

\bibitem{gough2001compiling}
J.~J. Gough and K.~J. Gough.
\newblock {\em Compiling for the. Net Common Language Runtime}.
\newblock Prentice Hall PTR, 2001.

\bibitem{grimmer2018cross}
M.~Grimmer, R.~Schatz, C.~Seaton, T.~W{\"u}rthinger, M.~Luj{\'a}n, and
  H.~M{\"o}ssenb{\"o}ck.
\newblock Cross-language interoperability in a multi-language runtime.
\newblock {\em ACM Transactions on Programming Languages and Systems (TOPLAS)},
  40(2):8, 2018.

\bibitem{ireland2009classification}
C.~Ireland, D.~Bowers, M.~Newton, and K.~Waugh.
\newblock A classification of object-relational impedance mismatch.
\newblock In {\em 2009 First International Confernce on Advances in Databases,
  Knowledge, and Data Applications}, pages 36--43. IEEE, 2009.

\bibitem{lee2003open}
J.~Lee and B.~Ware.
\newblock {\em Open Source Web Development with LAMP: Using Linux, Apache,
  MySQL, Perl, and PHP}.
\newblock Addison-Wesley Professional, 2003.

\bibitem{lerdorf2002programming}
R.~Lerdorf, K.~Tatroe, B.~Kaehms, and R.~McGredy.
\newblock {\em Programming Php}.
\newblock " O'Reilly Media, Inc.", 2002.

\bibitem{li2013jvm}
W.~H. Li, D.~R. White, and J.~Singer.
\newblock Jvm-hosted languages: they talk the talk, but do they walk the walk?
\newblock In {\em Proceedings of the 2013 International Conference on
  Principles and Practices of Programming on the Java Platform: Virtual
  Machines, Languages, and Tools}, pages 101--112. ACM, 2013.

\bibitem{lloyd1994practical}
J.~W. Lloyd.
\newblock Practical advtanages of declarative programming.
\newblock In {\em GULP-PRODE (1)}, pages 18--30, 1994.

\bibitem{milligan2017interactive}
M.~Milligan.
\newblock Interactive hpc gateways with jupyter and jupyterhub.
\newblock In {\em Proceedings of the Practice and Experience in Advanced
  Research Computing 2017 on Sustainability, Success and Impact}, page~63. ACM,
  2017.

\bibitem{northwood2018full}
C.~Northwood.
\newblock {\em The Full Stack Developer: Your Essential Guide to the Everyday
  Skills Expected of a Modern Full Stack Web Developer}.
\newblock Springer, 2018.

\bibitem{ristic2005apache}
I.~Ristic.
\newblock {\em Apache security}.
\newblock O'Reilly Media, 2005.

\bibitem{rule2018exploration}
A.~Rule, A.~Tabard, and J.~D. Hollan.
\newblock Exploration and explanation in computational notebooks.
\newblock In {\em Proceedings of the 2018 CHI Conference on Human Factors in
  Computing Systems}, page~32. ACM, 2018.

\bibitem{serrano2006hop}
M.~Serrano, E.~Gallesio, and F.~Loitsch.
\newblock Hop: a language for programming the web 2. 0.
\newblock In {\em OOPSLA Companion}, pages 975--985, 2006.

\bibitem{zonca2018deploying}
A.~Zonca and R.~S. Sinkovits.
\newblock Deploying jupyter notebooks at scale on xsede resources for science
  gateways and workshops.
\newblock In {\em Proceedings of the Practice and Experience on Advanced
  Research Computing}, pages 1--7. 2018.

\end{thebibliography}
\end{document}